\documentstyle[hip-prep]{article}  

\volnumber{1}  \edyear{1997}  

\def\rightmark{Chiral Dynamics with Quark Degrees of Freedom}
\def\leftmark{Zhonghan Feng, et al.}

\title{Chiral Dynamics with Quark Degrees of Freedom}
\authors{
{\twerm Zhonghan Feng$^1$, D\'enes Moln\'ar$^1$ and L\'aszl\'o P\'al 
Csernai$^{1,2}$ %
}\\[2.812mm]
{\normalsize
\hspace*{-8pt}$^1$ Physics Department,Theoretical Physics Section \\
University of Bergen, All\'egaten 55, N-5007 Bergen, NORWAY\\[0.2ex]
\hspace*{-8pt}$^2$ Theoretical Physics Institute, University of Minnesota\\
116 Church Street SE, Minneapolis, Minnesota 55455, USA
}}
\abstract { Possibility to detect DCC fluctuations is discussed. It is
shown that interactions with quark background and dissipative effects
due to interactions in the chiral field may result in damping of
fluctuations. Since the magnitude of fluctuations depends strongly
on the initial state and speed of chiral phase transition accurate
evaluation of all modifying processes is required to predict
observability of DCCs.}

\begin{document}

\maketitle

\input{epsf.sty}

\vskip 1.5truecm 

The possiblity of producing
quark-gluon plasma in relativistic heavy ion collisons
is  exciting  especially from the point of view of
observing the chiral and de-confinement phase transitions.
Rajagopal and Wilczek have suggested
that if  chiral restoration 
 is of second order, nonequilibrium dynamics can generate
transient domains in which  macroscopic pion fields
 develop. As the chiral
field relaxes to the true vaccum in such a domain,
 it may lead to coherent
emission of pions \cite{rw93}. This kind of phenomenon is called
 a disoriented chiral
condensate (DCC). Bjorken and others
 pointed out that  DCC can lead to fluctuations
 in the charged and neutral pion spectra \cite{Bjo}.  
If we can observe this  phenomenon 
we can say the quark-gluon plasma has been produced, and
we see the signs of its hadronization.

The possibility  to resolve fluctuations of neutral
and charged pions critically depends on the size and energy content of these
domains. If the domains are roughly pion sized, the effect of DCC is too small
to be resolved in experiments. If domains are large enough, the effects of DCC
can have measurable consequences. All these signals:
the isospin fluctuations, the enhancement of the pion spectrum at low $p_T$,
and the suppression of HBT correlations
are characteristics of any large coherent source \cite{sg95}.

Most dynamical studies \cite{91,ks94,ggp94,gm94,hw94}
 of  DCC have been carried out
within the framework of the linear sigma model.
Using this model, one can study
the possibility of DCC formation and the domain size  of DCC.

The Lagrangian density of the linear sigma model can be written as :
\begin{equation} \label{Lag}
{\cal L} = {\cal L}_q + {\cal L}_{SM} =
{\cal L}_q 
        + \frac{1}{2} \left(\partial_\mu\sigma\ \partial^\mu\sigma +
          \partial_\mu{\vec{\pi}}\ \partial^\mu{\vec{\pi}}\right)
         -U(\sigma, {\vec{\pi}}),
\end{equation}
where
\begin{equation} \label{pot}
U(\sigma, {\vec{\pi}})= \frac{\lambda}{4}
\left(\sigma^2+{\vec{\pi}}^2-v^2\right)^2 - H \sigma
+  \left( m_\pi^2 f_\pi^2 - \frac{m_\pi^4}{4 \lambda} \right),
\end{equation}
is the so-called Mexican Hat potential,
and ${\cal L}_q$  is the contribution of quarks,
\begin{equation}
{\cal L}_q =
         \bar{q} \left[i\gamma_\mu \partial^\mu -
         g (\sigma + i \gamma_5 {\vec{\tau}} {\vec{\pi}})  \right] q.
\end{equation}
Here 
$q$ stands for the light quark fields,  $(u, d)$,  while $\sigma$ and
${\vec{\pi}}=(\pi_1, \pi_2, \pi_3)$ are the scalar meson and vector meson
fields which together form a 4-component chiral field
$\Phi = (\sigma, \vec{\pi})$. The last constant term is included to secure
that the minimum of the potential is $U_{min}=0$. 
 In the {\em linear} sigma model, the
fields $\sigma$ and $\vec{\pi}$ are treated independently without any
constraint, and coupled to the fermion field as in the above equation.
If we impose the constraint
$ \sigma^2+\vec{\pi}^2=f^2_\pi$,  we will have the {\em nonliner}
sigma model.
Without the term $H\sigma$  this
Lagrangian is invariant with respect to the $SU_{L}(2)\otimes SU_{R}(2)$
chiral transformations.

The parameters in this Lagrangian are chosen in such a way \cite{cm95} 
that in {\em normal vacuum} at $T=0$ chiral symmetry is spontaneously
 broken and expectation values of the meson fields are
\begin{equation} \label{nvac}
\langle \sigma \rangle=f_{\pi},~~\langle{\vec{\pi}}\rangle=0, 
\end{equation}
where $f_{\pi}$=93 MeV is the pion decay constant.  In the vacuum with
broken chiral symmetry pions represent soft ``azimuthal'' excitations of
the chiral field.  To have the correct pion mass in vacuum, $m_{\pi}$=138
MeV, one should take
\begin{equation} \label{par}
v^2=f_{\pi}^2-\frac{m_{\pi}^2}{\lambda},\ \ \ \ \ H=f_{\pi}m_{\pi}^2.
\end{equation}
The parameter $\lambda$ is related to the sigma mass, 
$m_{\sigma}^2 = 2\lambda f_{\pi}^2+m_{\pi}^2$, which can be chosen to be
about 0.6 GeV (then $\lambda \approx 20$).  Sigmas represent stiff,
``radial'' excitations of the chiral field.  The remaining coupling
constant $g$ can be fixed by the requirement that the effective quark
mass, $(m_q^2 = g (\sigma^2 + \vec{\pi}\/^2))$, 
in broken vacuum, $m_q=g\langle \sigma \rangle=gf_{\pi}$,
coincides with the constituent quark mass in hadrons, about 1/3 of the
nucleon mass $m=m_N/3$.  This gives $g\approx (m_N/3)/f_{\pi}\approx 3.3$.

When we take the mean-field approximation, ignore all loop contributions and
consider $\sigma$ and $\vec{\pi}$ as classical fields, the equation of motion
of chiral field can be written as :
\begin{eqnarray}
\partial_\mu \partial^\mu\sigma(x) + \lambda \left[\sigma^2(x)+
{\vec{\pi}}^2(x) - v^2  \right] \sigma(x)& - H
&=\ - g \rho_S(x),
\nonumber \\
\partial_\mu \partial^\mu\vec{\pi}(x) + \lambda \left[\sigma^2(x)+
{\vec{\pi}}^2(x) - v^2  \right] \vec{\pi}(x)&
&=\ - g \rho_P(x).
\nonumber \\
\label{e001a}
\end{eqnarray}
Here $\rho_S=\langle \bar{q} q\rangle $ and $\rho_P=i\langle \bar{q}\gamma_5
\vec{\tau} q\rangle$ are scalar and pseudoscalar quark densities,  which
should be determined self-consistently from the
motion of $q$ and $\bar{q}$ in background meson fields.
The scalar and pseudoscalar densities can be
represented as
\begin{equation}
\rho_S(x) = g a(x)\sigma(x), ~~~\rho_P(x) = g a(x)\vec{\pi}(x),
\label{den}
\end{equation}
where $a(x)$ is expressed in terms of the
momentum distribution of quarks and antiquarks $f(x, p)$,
$$
a(x) = \nu_q \int \frac{d^4p}{(2\pi\hbar)^3}\
2\delta(p^\mu p_\mu - m^2(x)) f(x, p)
$$
\begin{equation}
\longrightarrow\frac{\nu_q}{(2\pi\hbar)^3} \int \frac{d^3p}{E(x, {\bf p})}
[n_q(x, {\bf p}) + n_{\bar{q}}(x, {\bf p})] .
\label{e3}
\end{equation}
Here $\nu_q$ is the degeneracy factor of quarks.
Let us assume that the quark distribution is an ideal Ferimi
distribution at the freeze-out time $\tau_0$ ($\tau_0 \sim 7fm/c$)
and the subsequent distribution for time $\tau > \tau_0$ is given
by the solution of the Vlasov equation \cite{cm95}.
The integral of this source term $a(x)$ in the equation of motion
is evaluated  for the $\mu=0$,  $m_q=0$
case analytically earlier in ref.\cite{cm95}.  For $m_q\neq0$, 
$$
a(x)=\frac{2\nu_q}{(2\pi)^2\alpha}\frac{1}{\sqrt{1-\alpha^2}}
     \int_{0}^{\infty}
     \frac{q}{e^{\beta \sqrt{q^2+\alpha^2m_0^2}}+1}
     \arcsin \sqrt{\frac{1-\alpha^2}
     {1+\frac{\alpha^2 m^2}{q^2}}}dq,
$$
where $\alpha=\tau_0/\tau$, $m=m(\tau)$ and $m_0=m(\tau_0)$
are the quark masses at $\tau$ and $\tau_0$.
This integral can be numerically evaluated by using nine
point Laguerre integral formula.  This way we can follow the
solution of the equation of motion numerically from arbitrary
initial condition at $\tau_0$ during the whole symmetry breaking process.
In ref. \cite{cm95} only an estimate was given for the
initial growth rate of the instability.

Considering an $N$-dimensional uniform scaling expansion, ($N=0,1,2,3$)
the differential
operator,  $\partial_\mu \partial^\mu$,  in terms of the
variables,  proper time,  $\tau$,  and approproate N-dimensional
rapidity,  $\eta$,  takes the form
\begin{equation}
\left[ \partial_{\tau}^2 + \frac{N}{\tau} \partial_\tau
- \frac{1}{\tau^2} \partial_{\eta}^2 \right] ,
\end{equation}
where the term,  $- N \tau^{-1} \partial_\tau$, describes the damping caused
by the collective expansion of the system. A similar damping can occur due to
interactions,  particularly due to thermal interactions if our system is 
interacting with heat-bath.  To describe such 
damping we might add an additional damping term
which is present even in the absence of collective expansion ($N=0$):
$ - \gamma \partial_\tau$.

Assuming further that our initial condition satisfies the scaling
symmetry,  i.e.,  it depends only on $\tau$ but not on $\eta$,  such
an $\eta$ independence is conserved by the equations of motion. Thus for
the studies we describe in the following we drop the rapidity
dependent terms in the equations of motion.  The arising equations
in the $(\sigma,  \vec{\pi})$ coordinates are
\begin{eqnarray}
\partial_{\tau}^2              \sigma  \ =\
\left(
- \frac{N}{\tau} \partial_\tau
- \gamma_\sigma  \partial_\tau
- \lambda (\sigma^2+{\vec{\pi}}^2)
+ \lambda v^2
\right) \sigma &
+H
&- g \rho_S,
\nonumber \\
\partial_{\tau}^2          \vec{\pi} \ =\
\left(
- \frac{N}{\tau} \partial_\tau
- \gamma_\pi  \partial_\tau
- \lambda (\sigma^2+{\vec{\pi}}^2)
+ \lambda v^2
\right) \vec{\pi} &
&- g \rho_P.
\label{eq33}
\end{eqnarray}
In the above equation, $\gamma_\pi$ and $\gamma_\sigma$ are damping
paramters of $\pi$ and $\sigma$ fields.


In estimating thermal damping of small amplitude fluctuations we may take
advantage of the estimated viscosity coefficient of the quark-gluon plasma
\cite{bmpr90} which was used earlier in estimating thermal nucleation rates 
\cite{ck92}. These yield a damping of about
\begin{equation}
\gamma \approx  0.2 - 10\ \frac{\rm c}{\rm fm}
\ \ \ \ \ \Gamma = 50 - 2000\ {\rm MeV} \\
\end{equation}
for $T\approx 100 $ MeV.

\begin{table}[htb]
\begin{center}
\begin{tabular}{rrr}
\hline
$m_\sigma$   &    $a$    &   $b$   \\
MeV          &           &          \\
\hline
400          & 0.0307   & 3.706    \\
600          & 0.0526   & 3.446    \\
800          & 0.0700   & 3.547    \\
\hline
\end{tabular}
\end{center}
\vspace{-4mm}
\caption[]{Fit parameters $a$ and $b$ to estimate the pion-width
as given in eq. (\ref{gammat}), for three sigma masses including
contributions of one- and two-loop self-energy diagrams, based on
ref. \cite{dlr94}. }
\end{table}

In finite temperature field theory we can estimate the damping of
a field using the imaginary-time propagators.  The damping or width is
characterized by the imaginary part of the self-energy.
These hadronic thermal widths should be interpreted as the damping
coefficients of wave packets propagating through a dispersive medium.

In the linear $\sigma$-model one-vertex   one-loop graphs do not 
contribute an imaginary part to the self energy \cite{ka89} thus the 
first contributions are given by two-vertex one-loop diagrams \cite{we83}. 
The damping of the $\pi$-field was evaluated in \cite{dlr94} by
including a  two-vertex one-loop self-energy graph with a dressed
sigma-meson propagator, plus two two-loop graphs and two two-loop
four-vertex graph contributions to the pion self-energy.

The resulting pion width can be approximated with
\begin{equation}
\Gamma_\pi = (a T)^b
\label{gammat}
\end{equation}
where $\Gamma_\pi$ and $T$ are measured in [MeV] and the parameters
$a$ and $b$ are given in Table 1.

For $T=100$ MeV and $m_\sigma = 600$ MeV, this yields $\Gamma_\pi
\approx 300$ MeV, which is comparable with the value above 
based on analogies of damping of waves in a viscous fluid.


At the last moment before quark freeze-out we have a thermal distribution
both for the quarks and the background fields. This moment was
estimated in ref.\cite{cm95} to be around $\tau_0 = 7$ fm 
and $T_0\approx 130$ MeV.
As an illustration we used the same initial condition as ref. \cite{hw94}
in our studies presented here.
\begin{eqnarray}
 \sigma(\tau_0)=0 &,&~~~\vec{\pi}(\tau_0)=0,\nonumber \\
\dot{\sigma}(\tau_0)= 1 {\rm MeV/fm} &,& ~~~
\dot{\pi_1}(\tau_0)= 5 {\rm MeV/fm}.
\label{ini}
\end{eqnarray}

\epsfysize=5.5cm
\begin{figure}[htb]
\vspace*{-0.4cm}
\center
\leavevmode
\epsfbox{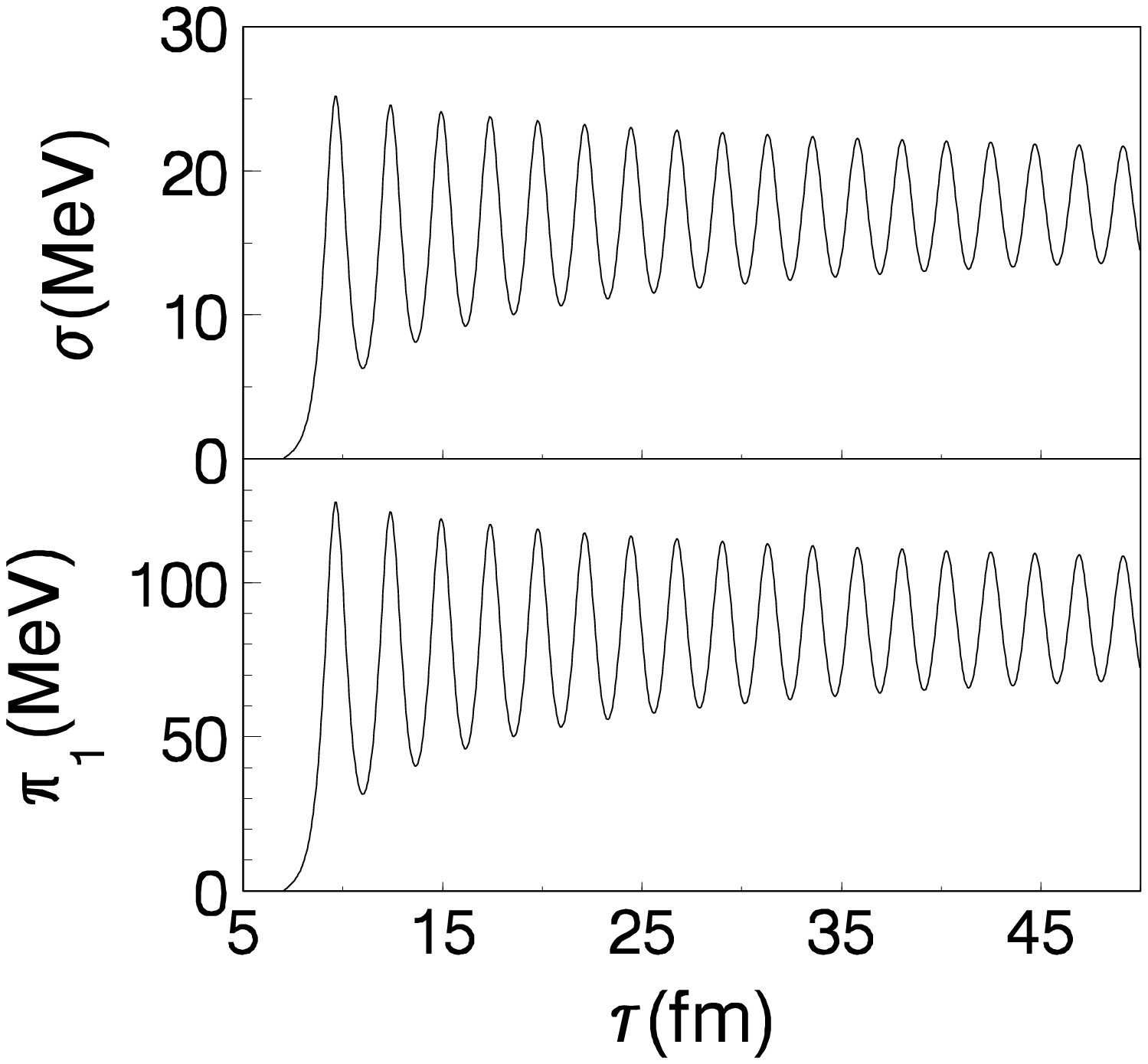}
       \vspace*{-1.0cm}
\caption{Proper time evolution of $\sigma$ and $\pi_1$ fields for
$H=0$, without quark-source terms and without damping.}
       \label{fig:fps1}
\end{figure}
Note that this initial condition is purely radial excitation
of $\Phi$ from the origin and thus it leads to correlated
$\sigma$ and $\vec{\pi}$ fluctuations.

\epsfysize=5.5cm
\begin{figure}[htb]
\vspace*{-1.cm}
\center
\leavevmode
\epsfbox{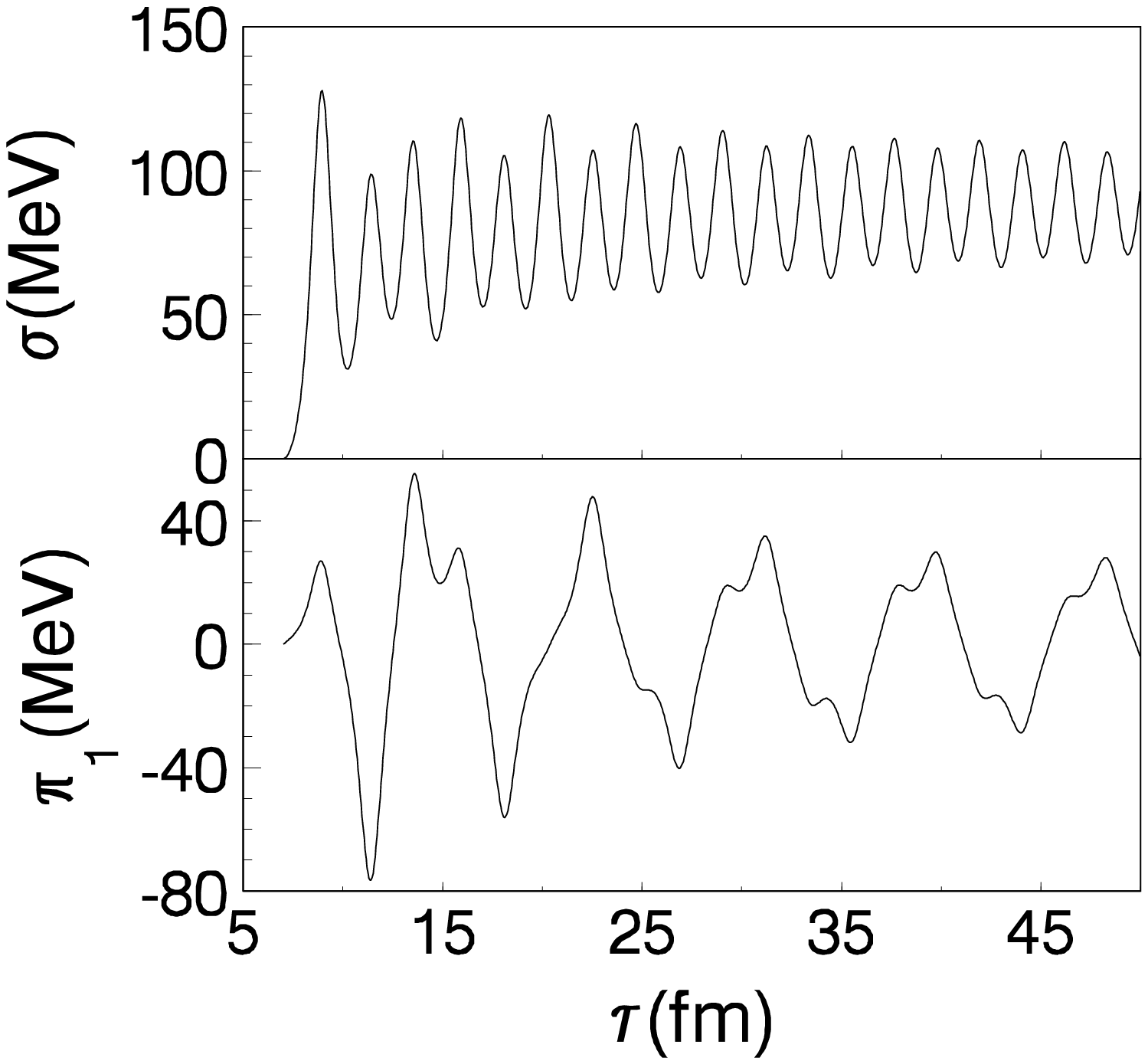}
       \vspace*{-1.0cm}
\caption{Proper time evolution of $\sigma$ and $\pi_1$ fields for
$H\neq0$, without quark-source terms and without damping.}
       \label{fig:fps2}
\end{figure}

\epsfysize=5.5cm
\begin{figure}[htb]
\vspace{-0.4cm}
\center
\leavevmode
\epsfbox{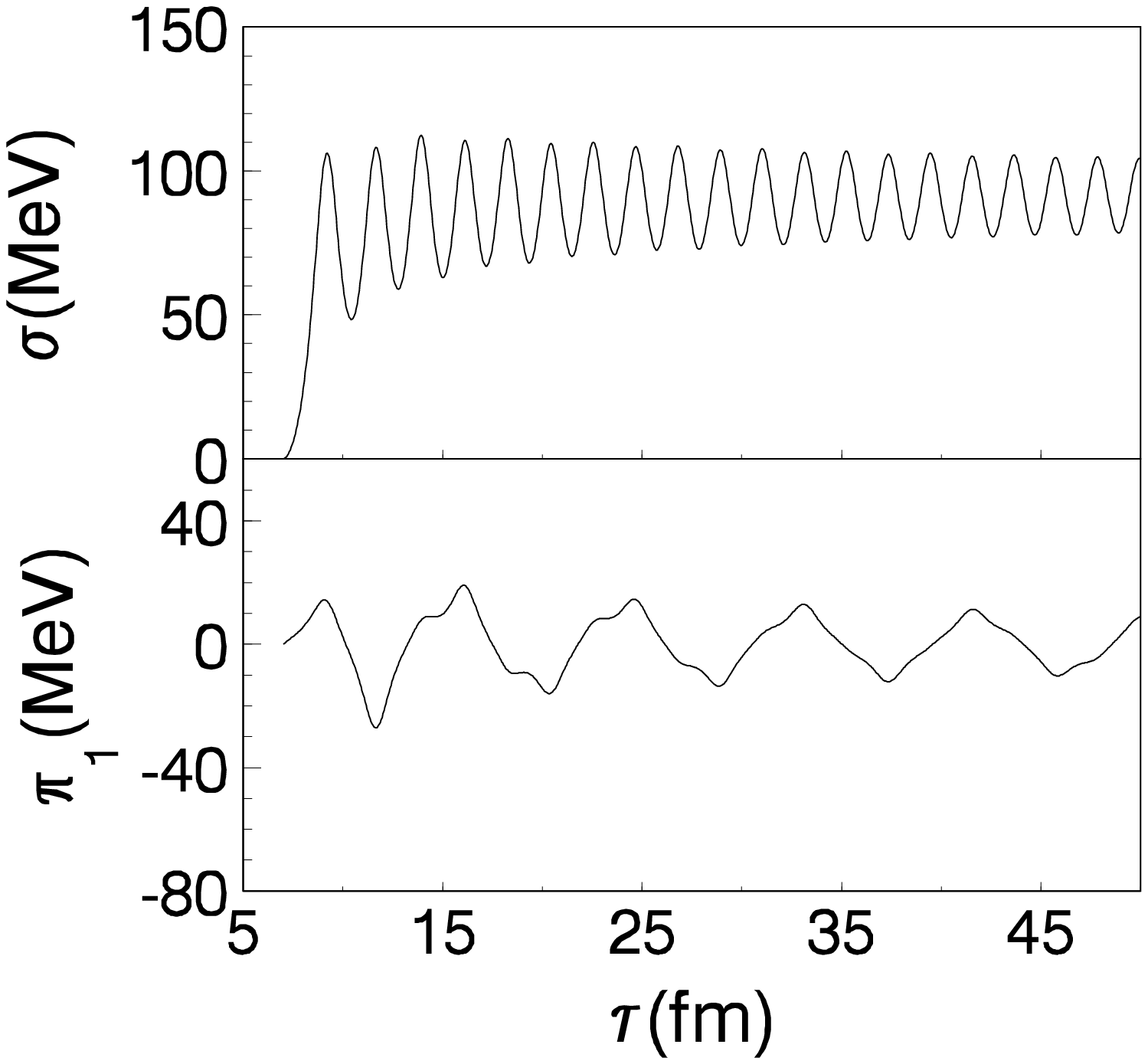}
       \vspace*{-1.0cm}
\caption{Proper time evolution of $\sigma$ and $\pi_1$ fields for
$H\neq0$, with quark-source terms but without damping.}
       \label{fig:fps3}
\end{figure}

\epsfysize=5.5cm
\begin{figure}[htb]
\vspace*{-1.cm}
\center
\leavevmode
\epsfbox{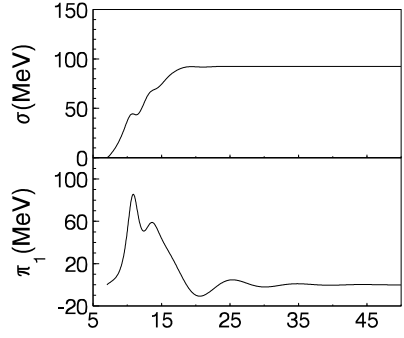}
       \vspace*{-1.0cm}
\caption{Proper time evolution of $\sigma$ and $\pi_1$ fields for
$H\neq0$, with quark-source terms and with damping,
$\gamma_\sigma=5.0 {\rm c/fm},\gamma_\pi=0.3 {\rm c/fm}$ .}
       \label{fig:fps4}
\end{figure}

For the above initial condition, we calaulate the proper time
evolution of sigma and pion fields. The features of different
scenarios are shown in Fig. \ref{fig:fps1}, Fig. \ref{fig:fps2},
Fig. \ref{fig:fps3}, and Fig. \ref{fig:fps4}.

In Fig. \ref{fig:fps1}, we solve the equation of motion without
symmetry breaking term, $H$, and do not consider the effect of
quark source
terms and damping. In this case, we exclude the background field.
The sigma and pion fields do not couple to the quark field and
they have complete symmetry. 
The oscillating frequencies of sigma and pion
fields are exactly  the same. They oscillate around different values
due to different initial conditions and the motion is radial in the
$(\sigma,\pi_1)$ plane. 

In Fig. \ref{fig:fps2}, we include the
symmetry breaking term, $H$.
 The dynamics is quite different compared
with Fig. \ref{fig:fps1}. In Fig. \ref{fig:fps2} the sigma field
oscillates nonlinearly around $f_\pi$, and the pion field oscillates 
around zero, their frequencies are also different. The sigma field takes
about 2 fm proper time to grow from zero to reach the true vacuum
expectation value $f_\pi$. On the other hand the pion field 
eventually tends to zero when the proper time gets large enough.

Compared to Fig. \ref{fig:fps2}, we include the quark source terms in
Fig. \ref{fig:fps3}. The evolution of sigma field does not change
too much and the oscillating amplitude of pion field is smaller
due to the source term. 

If $H\neq 0$ the transition is gradual, but fast. The source term leads
to a modification of the dynamics. Compared to the earlier dynamical
studies with a quenched initial condition, however, the difference
is not large and we get almost the same rapid rollover transition.

In Fig. \ref{fig:fps4}, we include the thermal damping term.
The sigma field  takes a longer time to grow from zero to
the true vacuum, then quickly tends to 
$f_\pi$ almost without oscillation. The fluctuation of
 the pion field is damped out quickly. It takes about 20 fm proper
time for the pion field to get to zero.  For the observability
it is important that  the fluctuations are rapidly damped out.

If we add the thermal damping term in the equation of motion,
we can see that the decrease of the amplitude of field oscillation
is very fast. It is a question if this later one is a
realistic approach, because  for small systems both
damping and fluctuations should be considered simulataneously.

\section*{Acknowledgments}
Discussions and suggestions for improvements are thanked to J. Randrup.

\section*{Notes} 
\begin{notes}
\item[a]
E-mail addresses:\\ 
Z.H. Feng: feng@kvark.fi.uib.no,\\
D. Molnar: molnard@kvark.fi.uib.no,\\
L.P. Csernai: csernai@fi.uib.no
\end{notes}

\hfill {\it Dedicated to Prof. K\'aroly Nagy on his 70th birthday} 

\bibliography{dccp}
\bibliographystyle{unsrt}
\end{document}